\documentclass[%
 reprint,
 amsmath,amssymb,
 aps,
]{revtex4-1}

\usepackage{graphicx}
\usepackage{dcolumn}
\usepackage{bm}
\usepackage{array}
\usepackage[all,dvips]{xy}

\begin{document}

\preprint{APS/123-QED}

\title{Solitary states for coupled oscillators}

\author{Patrycja Jaros$^1$,  Serhiy Brezetsky$^1$, Roman Levchenko$^3$, Dawid Dudkowski$^1$,Tomasz Kapitaniak$^1$, and Yuri Maistrenko$^{1,2}$}

\affiliation{$^{1}$Division of Dynamics, Technical University of Lodz, Stefanowskiego 1/15, 90-924 Lodz, Poland}
\affiliation{$^{2}$Institute of Mathematics and Centre for Medical and Biotechnical Research, National Academy of Sciences of Ukraine, Tereshchenkivska St. 3, 01030, Kyiv, Ukraine}
\affiliation{$^{3}$Taras Shevchenko National University of Kyiv, Volodymyrska St. 60, 01030 Kyiv, Ukraine}

\begin{abstract}
We demonstrate that \textit{solitary states} can be widely observed for networks of coupled oscillators with local, non--local and global couplings, and they preserve in both thermodynamic and Hamiltonian limits. We show that depending on units' and network's parameters, different types of solitary states occur, characterized by the number of isolated oscillators and the disposition in space. The creation of solitary states through the homoclinic bifurcation is described and the regions of co--existence of obtained states and typical examples of dynamics have been identified. Our analysis suggests that solitary states can be observed in a wide class of networks relevant to various real--world systems.
\vspace{\baselineskip}

\textit{Keywords}: solitary states, coupled oscillators, Kuramoto with inertia.
\end{abstract}

\maketitle

The discovery of chimera states in 2002 by Kuramoto and Battogtokh \cite{chim1} has opened a new chapter in nonlinear dynamics of coupled oscillators. Apart from well known synchronization problems \cite{synch} and chaotic dynamics \cite{chaos}, the area of interest of researchers has expanded onto new surprising kinds of structures, in which both coherent and incoherent types of behavior co--exist. Nowadays, the so--called chimeric patterns \cite{chim2} are thoroughly studied and their existence has been shown in different types of dynamical networks, both theoretically and experimentally, e.g. chemical \cite{chem1,chem2} and mechanical \cite{mech1,mech2,mech3} oscillators, electronic \cite{elec1,elec2} and laser \cite{laser1,ss2} systems. See \cite{more1,more2} for more references on the topic.

As with each new phenomenon, studies on chimera states allowed to introduce new and interesting ideas and describe many different concepts, based on the original one. An example of such a concept is the \textit{weak chimera state}, which has been defined by Ashwin and Burylko in 2015 \cite{weak1}. Weak chimeras arise in small networks of coupled oscillators and can be found in theoretical models \cite{weak1,weak2,weak3}, as well as experimental ones \cite{weak4}.

This Letter is about new weak chimera state patterns, i.e. solitary states. The term solitary comes from the Latin solitarius and can be understood as alone, lonely or isolated.
Indeed, a typical solitary state is identified as those including one or more isolated units oscillating with different mean frequency compare to the others frequency synchronized. Solitary states are different from the classical chimera states introduced in \cite{chim1} and \cite{chim2}, since the desynchronized units do not need to create a localized group as it is in the classical chimera states. Varying the initial conditions the desynchronized oscillators can arise at any place of the network space and in any combination. Due to this, the total number of solitary states in the network can grow exponentially with system size N illuminating eventually the phenomenon of spatial chaos \cite{sc1,sc2}.

In this Letter, we explain where the solitary states come from and pinpoint the conditions that allow them to exist. Previously, examples of solitary states have been found in different types of networks, e.g. Kuratomo model \cite{ss1}, delayed--feedback systems \cite{ss2}, in Stuart--Landau equations \cite{ss3} or in globally coupled identical oscillators with attractive and repulsive interactions \cite{ss4}. 
We give evidence that solitaty states are typical patterns on the route from regular dynamics to spatio--temporal chaos in the networks of coupled oscillators.

The Kuramoto--Sakaguchi model has been derived as an approximation of the Ginzburg-Landau equation \cite{chim1,KurSak1,KurSak2,KurSak3} and for now, it is a paradigmatic model to study general properties and nonlinear mechanisms in oscillatory networks of different nature. A number of variants of Kuramoto model has been proposed getting extensive applications, for example, in biology \cite{ref102} or neuroscience \cite{ref103,ref104,ref105}.

In our research we consider a network of coupled Kuramoto with inertia oscillators \cite{ss1,KurIn1,KurIn2,KurIn3,KurIn4}, which has been inspired by the  Ermentrout's study \cite{ref108}. For the first time this model has been analyzed by Tanaka, Lichtenberg and Oishi \cite{ref106,ref107}. Even earlier, in 1995, Antoni and Ruffo introduced Kuramoto model with inertia in Hamiltonian limit, i.e., without damping term,  as a network of classical rotators, which can also be interpreted as a mean-field XY Heisenberg model \cite{ref109}. Nowadays, Kuramoto with inertia is considered as the most appropriate model of power grids,  investigation of which represents an important practical problem \cite{Power1,Power2,Power3}. Another significant application is Josephson junctions \cite{Josephson}. An extended description of the origin of the Kuramoto model with inertia and its application can be found in \cite{KurIn2}.

Let us consider the following system of coupled identical oscillators

\begin{equation}
m {\ddot{\varphi}}_{i} + \varepsilon {\dot{\varphi}}_{i} = \frac{\mu}{2P+1} \sum\limits_{j=i-P}^{i+P} \sin ({\varphi}_j-{\varphi}_i-\alpha),
\label{KurInertia}
\end{equation}
where $i = 1, \ldots, N$ and $N$ is the size of the network. $\varphi_i$ describes the phase of $i$-th oscillator, while $m, \varepsilon, \alpha$ denote the mass, damping coefficient and phase lag, respectively. Each unit is connected with $P$ nearest neighbors to the left and to the right with equal coupling strength $\mu$.

\begin{figure}
\includegraphics[scale=0.42]{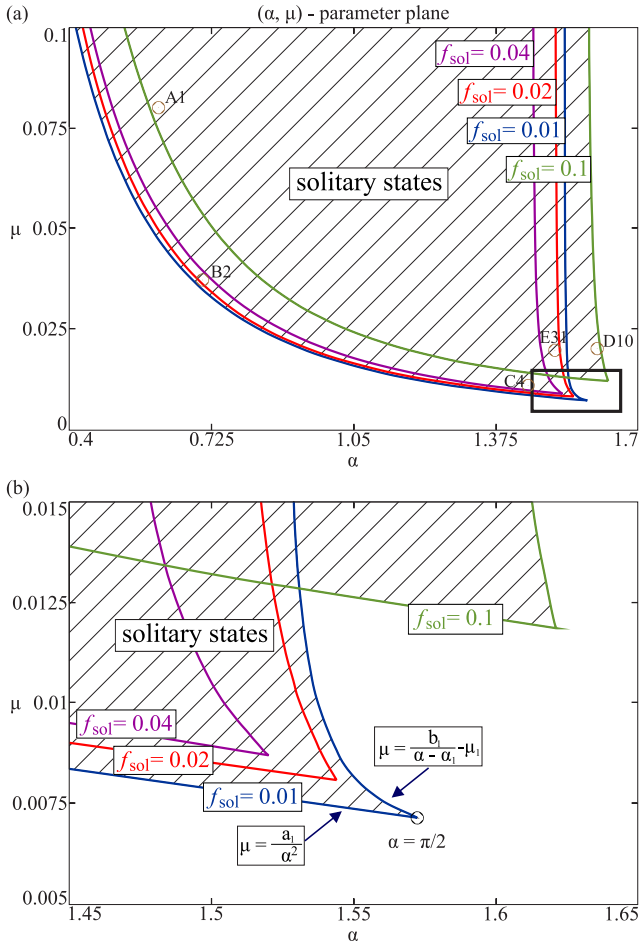}
\caption{(color online). In (a) the regions of co--existence of different types of solitary states in two parameters plane $(\alpha, \mu)$ for network (\ref{KurInertia}) with $P=40$ and $N=100$ are shown. Boundary of each region is denoted by a different color, namely blue, red, violet and green (states with $1,2,4$ and $10$ isolated oscillators, respectively). Points A1, B2, C4, D10 and E31 correspond to examples shown in Fig.~\ref{fig2}. In (b) the enlargement of black rectangle region marked in (a) is presented.}
\label{fig1}
\end{figure}

\begin{figure}
\includegraphics[scale=0.38]{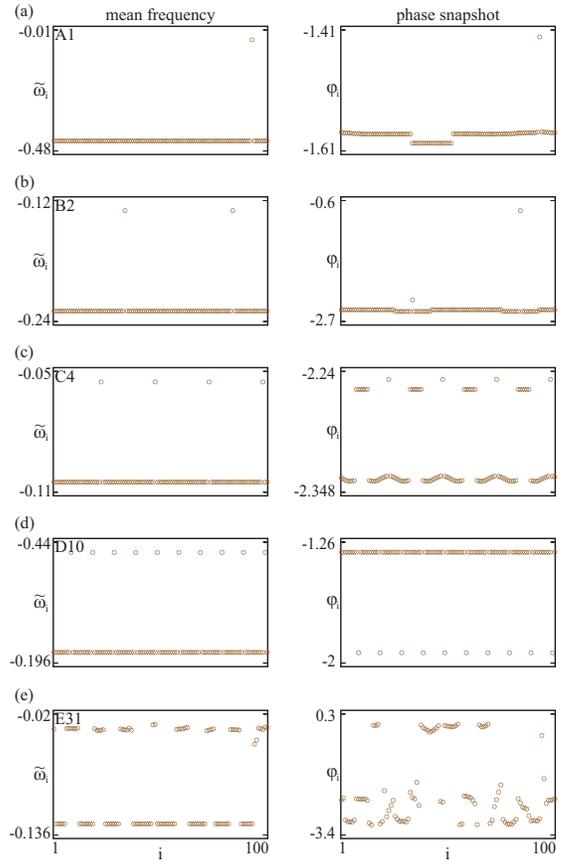}
\caption{(color online). Typical examples of solitary states observed for network (\ref{KurInertia}). Counting from the top (a) to the bottom (e), states correspond to parameters' values marked in Fig.~\ref{fig1} by A1, B2, C4, D10 and E31 points, respectively. In the left panel of each subfigure mean frequency snapshots are presented, while in the right one the phase snapshots are shown.}
\label{fig2}
\end{figure}

In Figs.~\ref{fig1}--\ref{fig2} we present the results of our studies on network (\ref{KurInertia}) for fixed parameters $m=1.0, \varepsilon=0.1, P=40$ and $N=100$ (in this case coupling is non--local, with coupling radius equal $r=0.4$).

The regions of appearance of solitary states types (characterized by the number of isolated oscillators) are shown in Fig.~\ref{fig1}(a). Results are presented in two parameters plane $(\alpha, \mu)$ and the boundary of each region is marked by a different color. To distinguish different types of states, the \textit{solitary fraction} $f_{sol} = \kappa / N$ is introduced and defined as the ratio between number of isolated units $\kappa \in \mathbb{N}$ and the size of the considered network $N$ (for clarity, the term solitary--$\kappa$ state will be used equivalently to $f_{sol}$ parameter). In Fig.~\ref{fig1}(b) one can see the enlargement of black rectangle region marked in Fig.~\ref{fig1}(a). On the other hand, in Fig.~\ref{fig2} one can observe typical examples of states discussed in Fig.~\ref{fig1}. Each subfigure Figs.~\ref{fig2}(a--e) (counting from the top to the bottom) corresponds to different parameters values, marked in Fig.~\ref{fig1} by A1, B2, C4, D10 and E31 points, respectively. Dynamics described in Fig.~\ref{fig2} is presented in the form of mean frequency (left panel) and phase snapshots (right panel).

We begin our study with the simplest, but not trivial situation of the only one isolated oscillator, i.e. solitary--1. The region of existence of this type of behavior is bounded in Fig.~\ref{fig1} by blue curves, while a typical example can be seen in Fig.~\ref{fig2}(a) for A1=$(0.6, 0.08)$. As one can see, the mean frequencies of all oscillators except one are equal (they are frequency synchronized), and a cluster pattern formed from these units is shown in the phase snapshot to the right.
It consists of two distinctive parts. Moreover, the phases of each part are also not identical and therefore, there are no phase synchronization inside the frequency synchronized cluster. This property relates to the considered non--local coupling scheme (parameter $P=40$ in Eq.~(\ref{KurInertia})) and vanishes for globally coupled network ($P=50$). 

Increasing the number of isolated oscillators, one can observe solitary--2 states, which example is shown in Fig.~\ref{fig2}(b) for parameters $\alpha=0.7, \mu=0.0372$ (point B2 located in red boundary region in Fig.~\ref{fig1}). In this case, mean frequency of two units is different, while the phase scenario is similar as previously (isolated units are oscillating independently from the main cluster, which oscillators are frequency synchronized).
In Fig.~\ref{fig2}(c) a solitary--4 state observed for C4=$(1.44, 0.011)$ (violet region boundaries in Fig.~\ref{fig1}) is shown. In this case the mean frequency profile is similar as in the previous examples, but the phase behavior within the frequency cluster is different: apart from  the isolated units, one can identify groups of coherent osillators where the phases of the oscillators are 'almost' equal.
With an increase of the number of solitaries, the behavior and observed scenarios becomes more involved. One such example for $\kappa=31$ solitaries is shown in Fig.~\ref{fig2}(e) for parameters $\alpha=1.5$ and $\mu=0.0197$.  Although there are still two mean frequency levels (as in the other examples), phases of the oscillators behave in a much more complicated way. 

In the study, we have observed many types of the solitary patterns, with the number of isolated oscillators up to half of the network size $N/2$ \cite{solregion}, with different space distribution and dynamics. We conclude that number of stable solitary states in model (\ref{KurInertia}) grows exponentially with $N$, giving rise to the phenomenon of spatial chaos \cite{spchaos}. One more interesting fact we observed is  that boundaries of the parameter  regions for different solitary states fit well to hyperbolic functions, i.e. power laws.  An example is shown for the blue boundary (solitary--1 state in Fig.~\ref{fig1}), where the left branch is described as $\mu=a_1 / {\alpha}^2$ for $a_1=0.01681$, while the right one is estimated by the curve $\mu=b_1 / (\alpha-{\alpha}_1) + {\mu}_1$, and the constants are aproximately found as $b_1=0.0000543459, {\alpha}_1=1.521524,$ and ${\mu}_1=0.0119794$.

\begin{figure}
\includegraphics[scale=0.49]{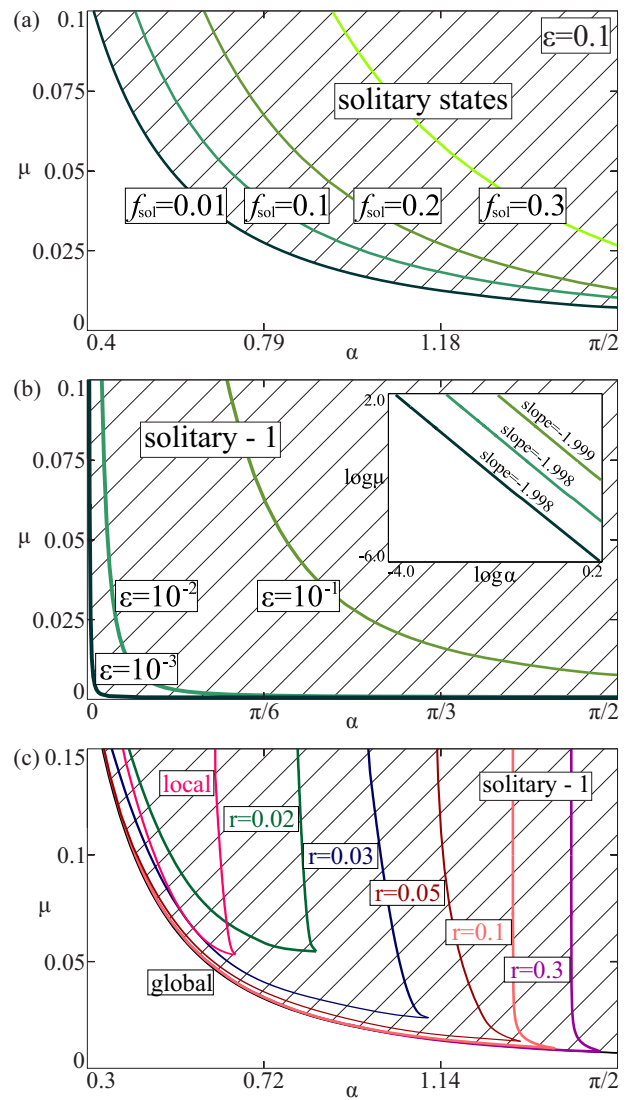}
\caption{(color online). In (a) the regions of co--existence of different types of solitary states in $(\alpha, \mu)$ plane for globally coupled network (\ref{KurInertia}) and $\varepsilon=0.1$ are shown for $f_{sol}=0.01,0.1,0.2,0.3$. Occurrence of solitary--1 states for varied damping coefficient (fixed $r=0.5$) and coupling radius (fixed $\varepsilon=0.1$) are presented in (b) and (c), respectively. Size $N=100$.}
\label{fig3}
\end{figure}

Results presented in Figs.~\ref{fig1}--\ref{fig2} have been obtained for Eq.~(\ref{KurInertia}) in the case of non--local coupling (with coupling radius $r=P/N=0.4$) and fixed damping $\varepsilon=0.1$. To our surprise, solitary states can also be observed for paradigmatic coupling schemes, i.e. globally, as well as locally coupled systems. The solitary states appearance for network (\ref{KurInertia}) with different types of coupling and various damping strength is presented in Fig.~\ref{fig3}.

In Fig.~\ref{fig3}(a) one can observe the bifurcation diagram for globally coupled network
\begin{equation}
m {\ddot{\varphi}}_{i} + \varepsilon {\dot{\varphi}}_{i} = \frac{\mu}{N} \sum\limits_{j=1}^{N} \sin ({\varphi}_j-{\varphi}_i-\alpha),
\label{KurGlobal} 
\end{equation}
where the variability of the solitary patterns is demonstrated in terms of {\it solitary fraction} $f_{sol} = \kappa / N$ ($\kappa$ is the number of isolated oscillators in the state). Parameter region for the solitary states is marked as hatched,  and boundaries for particular states with $f_{sol}=0.01,0.1,0.2,0.3$ are denoted by different colors. As one can see, the shape of the curves is similar to the one observed for the non--local model in Fig.~\ref{fig1}. Indeed, the dependence between $\mu$ and $\alpha$ remains quadratic hyperbolic: $\mu=c / {\alpha}^2$, where $c$ is approximately equal $0.0169, 0.0253, 0.0451, 0.1013$ for $f_{sol}=0.01, 0.1, 0.2, 0.3$, respectively.
It should be noted that because of the global coupling scheme, all oscillators except the isolated ones are perfectly phase sychronized. The mechanics of the solitary states appearance for Eq.~(\ref{KurGlobal}) is similar to the phenomenon of smallest chimera states \cite{small1,small2}, where the system dynamics can be reduced to a flow at two--dimensional torus, as a stable manifold of the whole phase space. 

To study the influence of damping, we have traced the solitary--1 bifurcation curve as $\varepsilon$ decreases. The results are presented in Fig.~\ref{fig3}(b), where one can see that with a decrease of damping the solitary curve approaches the coordinate axes $\alpha=0$ and $\mu=0$. As before, the curves are in power law $\mu=c / {\alpha}^2$, where $ c\approx 0.0169,  0.00016947,  0.0000016864$ for $\varepsilon=0.1, 0.01, 0.001$, respectively (the approximation to the quadratic slope is shown in inset). In consequence, when $\varepsilon \rightarrow 0$, solitary states occupy more and more parameter space so that they can be eventually observed in the whole range $(\alpha, \mu) \in (0, \pi/2) \times\mathbb{R}_{+}$.

On the other hand, in Fig.~\ref{fig3}(c) the solitary region for the non--locally coupled model (\ref{KurInertia}) is examined for varied coupling radius $r$, from global ($r=0.5$) to local ($r=0.01$), at fixed damping coefficient $\varepsilon=0.1$. As one can see, with a decrease of $r$ solitary patterns are less observable since parameter borders of their occurrence shrink. A nice property should be noted, i.e. region calculated for $r=r^{*}$ includes all the regions obtained for $r<r^{*}$, which occurs for all coupling radii except of the peculiar case of local coupling. 

To investigate closely the mechanism of creation of solitary states,  we have studied in more details the dynamics of globally coupled network (\ref{KurGlobal}) in the case of a single isolated unit, i.e. solitary--1 state (solitary--$\kappa$ states with larger $\kappa$ can be studied in the same way).
\begin{figure}
\includegraphics[scale=0.47]{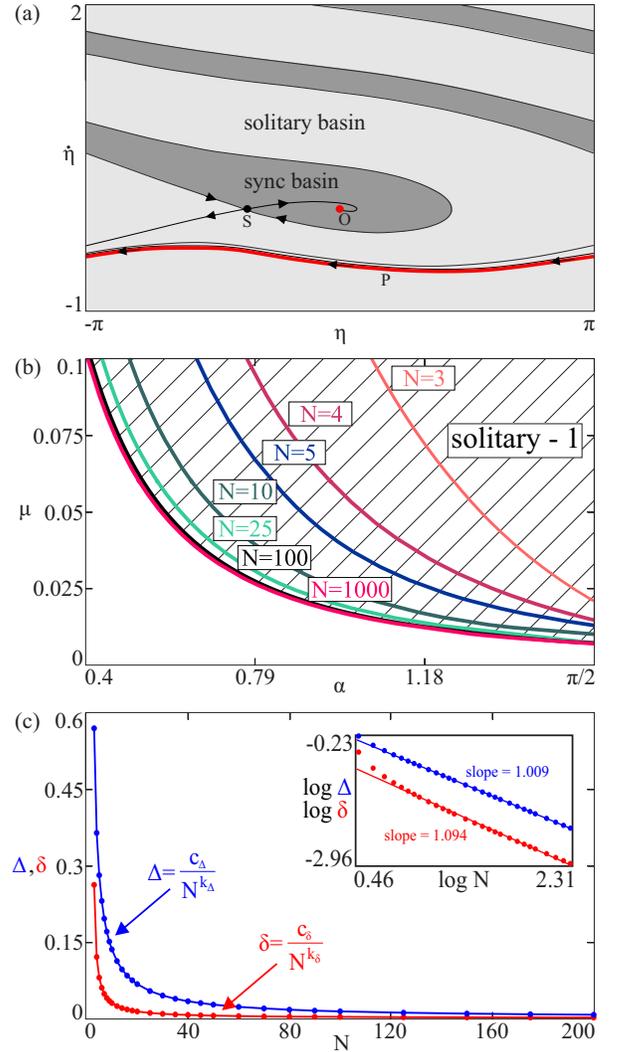}
\caption{(color online). In (a) the co--existence of synchronous and solitary states for simplified system (\ref{KurInertia3}) has been shown ($\alpha=1.0, \mu=0.06$ and $N=100$). Equilibrium O (sync basin) corresponds to the synchronization of all the units, while limit cycle P (solitary basin) appears as a result of homoclinic bifurcation and denotes solitary behavior. Regions of existence of solitary--1 states in $(\alpha, \mu)$ plane for varied size of network (\ref{KurInertia2}) have been presented in (b), where boundary of each region is denoted by a different color for $N=3,4,5,10,25,100,1000$. In (c) the influence of solitary--1 state on the dynamics of remaining oscillators is presented along with increase of network's size ($\alpha=1.0, \mu=0.15$). Parameters $\Delta$ and $\delta$ describe the amplitude of frequency fluctuations of synchronized group and the difference between mean frequency and $\Omega$, respectively. All subfigures correspond to the global coupling case and $\varepsilon=0.1$.}
\label{fig4}
\end{figure}
Without loss of generality let us assume that the first element is isolated and denote $\psi=\varphi_{1}$ and $\theta=\varphi_{2}, \ldots, \varphi_{N}$. In such a case, equation (\ref{KurGlobal}) simplifies as follows:
\begin{equation}
\begin{split}
m\ddot{\theta}+\varepsilon\dot{\theta}& =  -\dfrac{(N-1)\mu}{N}\sin(\alpha)+\dfrac{\mu}{N}\sin(\psi-\theta-\alpha), \\ 
m\ddot{\psi}+\varepsilon\dot{\psi}& =  \dfrac{(N-1)\mu}{N}\sin(\theta-\psi-\alpha)-\dfrac{\mu}{N}\sin(\alpha).
\label{KurInertia2}
\end{split}
\end{equation}
Introducing new state variable $\eta=\theta-\psi$, Eqs. (\ref{KurInertia2}) are reduced to 
\begin{equation}
\begin{split}
m\ddot{\eta}+\varepsilon\dot{\eta} = & -\dfrac{\mu}{N}((N-2)\sin(\alpha)+\sin(\eta+\alpha)+ \\
 & (N-1)\sin(\eta-\alpha)).
\label{KurInertia3}
\end{split}
\end{equation}

The dynamics of Eq. (\ref{KurInertia3}) for $\alpha=1.0, \mu=0.06$ and $N=100$ is presented in Fig.~\ref{fig4}(a). The system possesses two equilibria, i.e. stable origin O=$(0,0)$ (corresponding to the fully synchronized state) and saddle point S $\approx (-1.16,0)$ (marked as red and black dots, respectively). Critical points O and S co--exist with stable limit cycle P (red curve), which corresponds to the solitary state.  Orbit P appears as a result of a homoclinic bifurcation, which occurs when  stable and unstable manifolds of saddle S collide. After the bifurcation, one can identify the co--existence of two types of dynamics in Fig.~\ref{fig4}(a). Apart from the fully synchronous state (given by equilibrium O) a new type of behavior -- the solitary state (given by periodic orbit P) --  is born. Depending on the initial conditions in sync or solitary basins (shown in Fig.~\ref{fig4}(a)) that or another state will be realized in the model (\ref{KurInertia}). 

The role of the network's size $N$ in the solitary state occurence is illustrated in Fig.~\ref{fig4}(b), where solitary--1 bifurcation curves are shown for $N=3,4,5,10,25,100,1000$. As one can see, the smallest solitary region is observed at smallest system size $N=3$. With a slight increase of $N$  it rapidly expands, which is illustrated for $N=4,5,10$. However, further increase of $N$ does not change the solitary region essentially. Indeed, the curves for $N=100$ and $N=1000$ are almost indistinguishable, and the latter one can be assumed as a good approximation of the solitary--1 region boundary in the thermodynamic limit. 

To examine what is the influence of a single solitary oscillator on the synchronized group of remaining oscillators in the solitary --1 state, two additional parameters have been introduced, i.e. $\Delta = \max \dot{\theta} - \min \dot{\theta}$ and $\delta = \left|\Delta / 2 - \Omega \right|$, where $\Omega = -\frac{1}{\varepsilon} \mu \sin (\alpha)$ is the frequency of the fully synchronized state. The first parameter describes the amplitude of frequency fluctuations of oscillators in the frequency synchronized cluster, while the second one the difference between the mean frequency and $\Omega$ (both calculated after transient time). First, as it can be seen in Fig.~\ref{fig4}(c), for the smallest system size  $N = 3$ the values are rather large: $\Delta=0.56918$ and $\delta=0.26174$. Then, with an increase of $N$ to about 20 they both rapidly decrease and after, slowly converge to zero such that in the thermodynamic limit $N \to \infty$ both $\Delta$ and $\delta$ parameters equals zero. Furthemore, our numerical approximations provide the decay rate for both curves to be inversely proportional
to $N$ (in the simulations, the values for $N<6$ have been omitted):
 $\Delta=c_{\Delta} / N$, where $c_{\Delta}$=1.379141, while $\delta=c_{\delta} / N$, for which $c_{\delta}=0.368715$ (see Fig.~\ref{fig4}(c)).

In conclusion, we have shown that solitary states can be naturally found in networks of coupled oscillators with any coupling scenario, i.e. for locally, non--locally and globally coupled systems. Depending on system parameters,  different types of the discussed patterns appear, characterized mostly by the number of isolated units (but not only, as the space disposition of the units is also important in non--globally coupled case). Huge multistability of the solitary states is a generic property of the model leading to the phenomenon of spatial chaos (apparently, the number of such stable states grows exponentially with the system size $N$). It has been shown that solitary patterns are even more widely observable as the damping decreases to zero, possessing the whole parameter space in the Hamiltonian limit and furthermore, that their occurrence can be traced to the thermodynamic limit. We determine the mechanism of the solitary state appearance via a homoclinic bifurcation. We believe that the emergence of the solitary states is a universal phenomenon for networks of coupled oscillators of different nature.
\vspace{\baselineskip}

{\bf Acknowledgment}
This work has been supported by the Polish National Science Centre, MAESTRO Programme -- Project No 2013/08/A/ST8/00/780.


\begin{thebibliography}{70}
\bibitem{chim1}
Y. Kuramoto, and D. Battogtokh, Nonlinear Phenom. Complex Syst. 5, 380 (2002).

\bibitem{synch}
A. Pikovsky, M. Rosenblum, and J. Kurths, Synchronization: A Universal Concept in Nonlinear Sciences (Cambridge University Press, 2001).

\bibitem{chaos}
E. Ott, Chaos in Dynamical Systems (Cambridge University Press, 2012).

\bibitem{chim2}
D. M. Abrams, and S. H. Strogatz, Phys. Rev. Lett. 93, 174102 (2004).

\bibitem{chem1}
M. R. Tinsley, S. Nkomo, and K. Showalter, Nat. Phys. 8, 662 (2012).

\bibitem{chem2}
S. Nkomo, M. R. Tinsley, and K. Showalter, Phys. Rev. Lett. 110, 244102 (2013).

\bibitem{mech1}
E. A. Martens, S. Thutupalli, A. Fourriere, and O. Hallatschek, Proc. Natl. Acad. Sci. 110, 10563 (2013).

\bibitem{mech2}
T. Kapitaniak, P. Kuzma, J. Wojewoda, K. Czolczynski, and Y. Maistrenko, Sci. Rep. 4, 6379 (2014).

\bibitem{mech3}
L. Schmidt, K. Sch\"{o}nleber, K. Krischer, and V. Garcia--Morales, Chaos 24, 013102 (2014).

\bibitem{elec1}
L. V. Gambuzza, A. Buscarino, S. Chessari, L. Fortuna, R. Meucci, and M. Frasca, Phys. Rev. E 90, 032905 (2014).

\bibitem{elec2}
J. D. Hart, K. Bansal, T. E. Murphy, and R. Roy, Chaos 26, 094801 (2016).

\bibitem{laser1}
L. Larger, B. Penkovsky, and Y. Maistrenko, Nat. Commun. 6, 7752 (2015).

\bibitem{ss2}
V. Semenov, A. Zakharova, Y. Maistrenko, and E. Sch\"{o}ll, Europhys. Lett. 115, 10005 (2016).

\bibitem{more1}
M. J. Panaggio, and D. M. Abrams, Nonlinearity 28, R67 (2015).

\bibitem{more2}
E. Sch\"{o}ll, Eur. Phys. J. Special Topics 225, 891 (2016).

\bibitem{weak1}
P. Ashwin, and O. Burylko, Chaos 25, 013106 (2015).

\bibitem{weak2}
C. Bick, and P. Ashwin, Nonlinearity 29, 1468 (2016).

\bibitem{weak3}
Y. Suda, and K. Okuda, Phys. Rev. E 92, 060901(R) (2015).

\bibitem{weak4}
D. Dudkowski, J. Grabski, J. Wojewoda, P. Perlikowski, Y. Maistrenko, and T. Kapitaniak, Sci. Rep. 6, 29833 (2016).


\bibitem{sc1}
L. P. Nizhnik, I. L. Nizhnik, M. Hasler, Stable stationary solutions in reaction-diffusion systems consisting of a 1-d array of bistable cells, Int. J. Bif.Chaos 2002;12:261.

\bibitem{sc2}
I. Omelchenko, Y. Maistrenko, P. Hovel, E. Scholl, Loss of coherence in dynamical networks: Spatial chaos and chimera states, Phys Rev Lett 2011;106:234102.



\bibitem{ss1}
P. Jaros, Y. Maistrenko, and T. Kapitaniak, Phys. Rev. E 91, 022907 (2015).

\bibitem{ss3}
K. Premalatha, V. K. Chandrasekar, M. Senthilvelan, and M. Lakshmanan, Phys. Rev. E 94, 012311 (2016).

\bibitem{ss4}
Y. Maistrenko, B. Penkovsky, and M. Rosenblum, Phys. Rev. E 89, 060901(R) (2014).

\bibitem{KurIn1}
T. Bountis, V. G. Kanas, J. Hizanidis, and A. Bezerianos, Eur. Phys. J. Special Topics 223, 721 (2014).

\bibitem{KurIn2}
S. Olmi, A. Navas, S. Boccaletti, and A. Torcini, Phys. Rev. E 90, 042905 (2014).

\bibitem{KurIn3}
S. Olmi, E. A. Martens, S. Thutupalli, and A. Torcini, Phys. Rev. E 92, 030901(R) (2015).

\bibitem{KurIn4}
S. Olmi, Chaos 25, 123125 (2015).

\bibitem{KurSak1}
Y. Kuramoto, Chemical Oscillations, Waves and Turbulence (Springer, Berlin, 1984).

\bibitem{KurSak2}
H. Sakaguchi, and Y. Kuramoto, Prog. Theor. Phys. 76, 576 (1986).

\bibitem{KurSak3}
H. Sakaguchi, S. Shinomoto, Y. Kuramoto, Prog. Theor. Phys. 77, 1005 (1987).



\bibitem{ref102}

A. Pikovsky, M. Rosenblum, J. Kurths, Synchronization: A Universal Concept in Nonlinear Sciences, Cambridge University Press, 2001.

\bibitem{ref103}
M. Breakspear, S. Heitmann, A. Daffertshofer, Front Hum Neurosci. 4.

\bibitem{ref104}
 J. Cabral, H. Luckhoo, M. Woolrich, M. Joensson, H. Mohseni, A. Baker, M. Kringelbach, D. G., NeuroImage 90.

\bibitem{ref105}
 D. Cumin, C. P. Unsworth, Physica D. 226 (2).
 
\bibitem{ref108}
  B. Ermentrout, J. Math. Biol. 29, 571 (1991).
 
\bibitem{ref106}
 H.-A. Tanaka, A. J. Lichtenberg, and S. Oishi, Phys. Rev. Lett. 78, 2104 (1997).
 
\bibitem{ref107}
 H.-A. Tanaka, A. J. Lichtenberg, and S. Oishi, Physica D 100, 279 (1997).
 

 
\bibitem{ref109}
 M. Antoni, S. Ruffo, Phys. Rev. E 52, 2361 (1995).




\bibitem{Power1}
F. Salam, J. E. Marsden, and P. P. Varaiya, IEEE Trans. Circuits Sys. 31, 673 (1984).

\bibitem{Power2}
G. Filatrella, A. H. Nielsen, and N. F. Pedersen, Eur. Phys. J. B 61, 485 (2008).

\bibitem{Power3}
M. Rohden, A. Sorge, M. Timme, and D. Witthaut, Phys. Rev. Lett. 109, 064101 (2012).

\bibitem{Josephson}
B. R. Trees, V. Saranathan, and D. Stroud, Phys. Rev. E 71, 016215 (2005).

\bibitem{solregion}
For fixed parameters $\alpha$ and $\mu$ different types of solitary patterns may co--exist. E.g., the region of solitary--4 states is included in the region of solitary--2 ones, which in turn is included in the solitary--1 states area. However, for a higher number of isolated oscillators such inclusion scenario breaks down. E.g., the region of solitary--10 is located more to the right in Fig.~\ref{fig1} and it only intersects with previously calculated sets.  

\bibitem{spchaos}
I. Omelchenko, Y. L. Maistrenko, P. H\"{o}vel, and E. Sch\"{o}ll, Phys. Rev. Lett. 106, 234102 (2011).

\bibitem{small1}
Y. Maistrenko, S. Brezetsky, P. Jaros, R. Levchenko, and T. Kapitaniak, Phys. Rev. E 95, 010203(R) (2017).

\bibitem{small2}
J. Wojewoda, K. Czolczynski, Y. Maistrenko, and T. Kapitaniak, Sci. Rep. 6, 34329 (2016).
\end{thebibliography}
\end{document}